\documentclass{PoS}
\title{Recent results in QCD sum rule calculations of heavy meson properties}
\ShortTitle{Recent results in QCD sum rule calculations of heavy meson properties}
\author{\speaker{Dmitri Melikhov}\\
HEPHY, Austrian Academy of
Sciences, Nikolsdorfergasse 18, A-1050 Vienna, Austria\\ 
Faculty of Physics, University of Vienna, Boltzmanngasse 5, A-1090 Vienna, Austria\\ 
SINP, Moscow State University, 119991, Moscow, Russia\\
E-mail: \email{dmitri\_melikhov@gmx.de}}

\abstract{
This talk reviews the recent progress in the extraction of bound-state characteristics from 
the operator-product expansion (OPE) for field-theory correlators, which constitutes 
the basis of the method of QCD sum rules. This progress is mainly related to a deeper understanding 
of one of the key ingredients of the method of sum rules -- the effective continuum threshold. 
The latter determines 
to a great extent the numerical values of the bound-state parameters obtained by sum rules. 
The understanding of properties of the effective continuum threshold allows one to formulate 
a new algorithm for fixng this quantity for various correlators and in this way 
gain control over the systematic uncertainties of the bound-state parameters extracted from sum rules. 
We start with examples from quantum mechanics, where bound-state properties may be 
calculated independently in two ways: exactly, by solving the Schr\"odinger equation,
and approximately, by the method of dispersive sum rules. Knowing the exact solution is crucial as it allows 
us to control each step of the sum-rule extraction procedure. 
On the basis of this analysis, we formulate several improvements of the standard procedures adopted in the 
method of sum rules in QCD. We then show the impact of these modifications on the extraction of the 
decay constants of heavy mesons.}

\FullConference{The XXth International Workshop High Energy Physics and Quantum Field Theory     \\
          September 24 - October 1,  2011 \\
         Sochi, Russia}

\begin{document}
\section{Introduction}
The method of dispersive sum rules \cite{svz} is one of the widely used methods for 
obtaining properties of ground-state hadrons in QCD. The method involves two steps: 
\begin{itemize}
\item[(i)] One calculates the relevant correlator in QCD at relatively small values of the Eucledian time;  
\item[(ii)] One applies numerical procedures suggested by quark-hadron duality in order to isolate the ground-state 
contribution from this correlator. These numerical procedures cannot determine a single value of the 
ground-state parameter but should provide the band of values containing the true 
hadron parameter. This band is a systematic, or intrinsic, uncertainty of the method of sum rules. 
\end{itemize}
An unbiased judgement of the reliability of the extraction procedures adopted in the method of sum rules 
may be acquired by applying these procedures to problems where the ground-state parameters
may be found independently and exactly as soon as the parameters 
of theory are fixed. Presently, only quantum-mechanical potential models 
provide such a possibility: 
(i) the bound-state parameters (masses, wave functions, form factors) are known precisely from the Schr\"odinger equation;
(ii) direct analogues of the QCD correlators may be calculated exactly. 

Making use of these models, we studied the extraction of ground-state parameters 
from different types of correlators: namely, the ground-state decay 
constant from two-point vacuum-to-vacuum correlator \cite{lms_2ptsr}, the form factor from three-point 
vacuum-to-vacuum correlator \cite{lms_3ptsr}, and the form factor from vacuum-to-hadron correlator \cite{m_lcsr}. 
We have demonstrated that the standard procedures adopted in the method of sum rules not always work properly: 
the true value of the bound-state parameter was shown to lie outside 
the band obtained according to the standard criteria. 
These results gave us a solid ground to claim that also in QCD the actual accuracy of the method may be 
worse than the accuracy expected on the basis of applying the standard criteria. 

We realized that the main origin of these problems of the method originate from an over-simplified model 
for hadron continuum which is described as a perturbative contribution above a constant Borel-parameter 
independent effective continuum threshold. 
We introduced the notion of the {\it exact effective continuum threshold}, 
which corresponds to the true bound-state parameters: in potential models the true hadron 
parameters are known and the exact effective continuum thresholds for different correlators may be calculated. 
We have demonstrated that 
\begin{itemize}
\item[(i)]
The effective continuum threshold is not a universal quantity; it depends on the correlator considered 
(i.e., it is different for two-point and three-point vacuum-to-vacuum correlators); 
\item[(i)]
The effective continuum threshold depends on the Borel parameter and, for the form-factor case, 
also on the momentum transfer \cite{lms_3ptsr,m_lcsr,braguta,irinaPoS,irina}. 
In QCD, the effective threshold depends in addition on the renormalization scale. 
\end{itemize}
In the recent publications \cite{lms_new} we proposed a new algorithm for extracting 
the parameters of the ground state. 
The idea formulated in these papers is to relax the standard assumption of a Borel-parameter 
independent effective continuum threshold and to allow for a Borel-parameter dependent quantity. 
This talk explains the details of this procedure and its application to decay constants of heavy mesons. 


\section{\label{Sect:QM}OPE and sum rule in quantum-mechanical potential model}
\begin{figure}[!hb]
\begin{center}
\begin{tabular}{c}
\includegraphics[width=8cm]{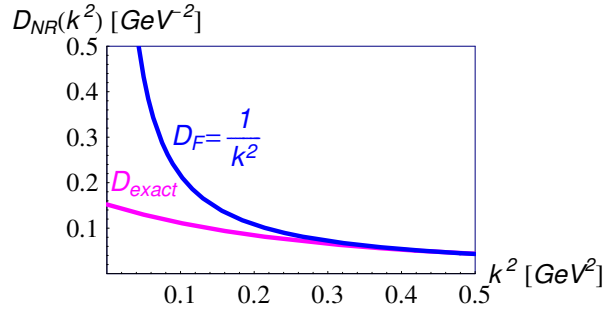}\\
\end{tabular}
\caption{\label{Plot:propagators}
Feynman vs exact propagator of a confined nonrelativistic particle.}
\end{center}
\end{figure}

Within the standard calculations based on perturbation theory, one start with a free Lagrangian and includes 
interactions order by order. Respectively, the basic object for perturbative calculations is the Feynman 
propagator. However, in a confining theory like QCD, the full propagator in the ``soft'' region of small momenta 
differs strongly from the Feynman propagator. The Feynman propagator of a nonrelativistic particle with the mass $m$ 
has the form 
\begin{eqnarray}
D_{\rm F}(E,\vec k^2)=\frac{1}{\vec k^2-2mE-{\rm i}0}
\end{eqnarray}
Fig.~\ref{Plot:propagators} compares the Feynman and the exact propagators at $E=0$ in a quantum-mechanical 
potential model for the case of the harmonic-oscillator potential $V=m\omega^2r^2/2$.  
This simple model allows one to calculate the exact propagator of a confined particle 
explicitly and illutrates the typical situation for a confining potential of a general form. 
Obviouly, the Feynman and the exact propagators are close to each other for large momentum transfers, 
but differ sizeably in the ``soft'' region. As soon as the ``soft'' momenta are essential in Feynman diagrams of the 
perturbation theory, nonperturbative effects in the propagators become essential. 
At large values of $\vec k^2$, one obtains 
\begin{eqnarray}
D_{\rm exact}(\vec k^2)=D_{\rm F}(\vec k^2)+c_2 \frac{\omega^2}{\vec k^2}+c_4 \frac{\omega^4}{\vec k^4}+\ldots
\end{eqnarray}
Nonperturbative effects in the propagator may be described in terms of an expansion in powers of $\omega^2/\vec k^2$. 
These nonperturbative effects appear then as power corrections in the correlation functions. 

Let us consider a realistic quantum-mechanical potential model containing both the Coulomb and the confining interactions. 
The corresponding Hamiltonian has the form 
\begin{eqnarray}
\label{conf+coulomb}
H=H_0+V(r); \qquad H_0=\frac{k^2}{2m}, \qquad V(r)=V_{\rm conf}(r)-\frac{\alpha}{r}.
\end{eqnarray}
Polarization operator $\Pi(E)$ and its Borel transform $\Pi(T)$ [$E\to T$, $1/(H-E)\to \exp(-H T)$, $T$ the Borel parameter] 
are defined via the full Green function $G(E)=1/(H-E)$ \cite{nsvz}: 
\begin{eqnarray}
\Pi(E)=\langle \vec r_f=0|\frac{1}{H-E}|\vec r_i=0\rangle,\qquad 
\Pi(T)=\langle \vec r_f=0|\exp(- H T)|\vec r_i=0\rangle.  
\end{eqnarray}
The expansion of $G(E)$ and $\Pi(E)$ in powers of the interaction is obtained with the help of 
the Lippmann-Schwinger equation 
\begin{eqnarray}
\label{lipp}
G(E)=G_0(E)-G_0(E)V G_0(E)+G_0(E)V G_0(E)V G_0(E)+\ldots,
\end{eqnarray}
where $G_0(E)=1/(H_0-E)$. Since the interaction contains now two parts, $V_{\rm conf}(r)$ and $\frac{\alpha}{r}$, 
the expansion (\ref{lipp}) is a double expansion in powers of $V_{\rm conf}$ and $\alpha$. 
E.g., for the case $V_{\rm conf}(r)=\frac{m\omega^2 r^2}2$ one easily obtains the corresponding 
double expansion in powers of $\alpha$ and $\omega T$ (see Fig.~\ref{Plot:Pi}): 
\begin{figure}[!b]
\begin{center}
\begin{tabular}{c}
\includegraphics[width=12cm]{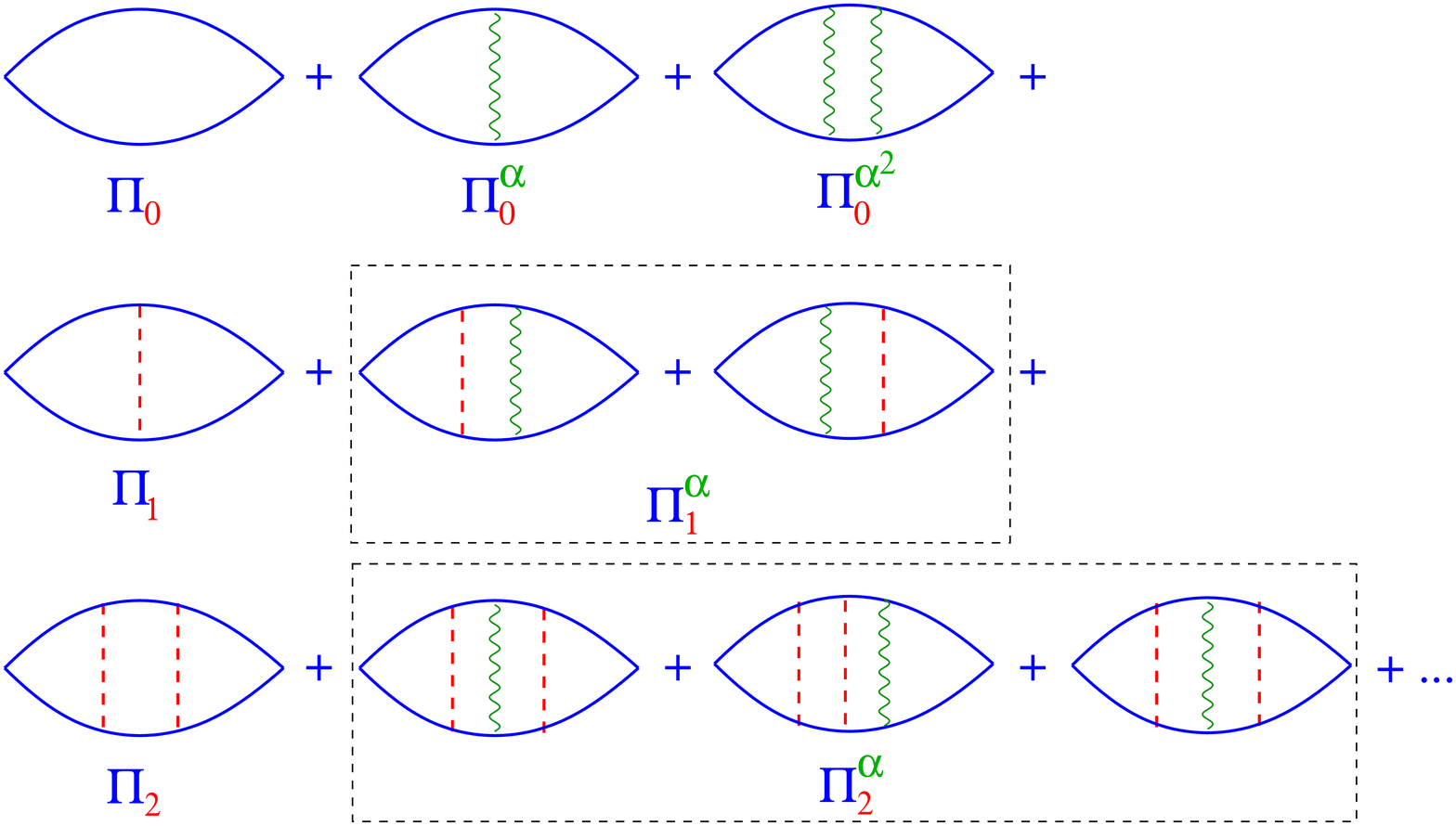}\\
\end{tabular}
\caption{\label{Plot:Pi}
Expansion of the polarization functions in powers of the interaction.}
\end{center}
\end{figure}

\begin{eqnarray}
\Pi_{\rm OPE}(T)&=&\Pi_{\rm pert}(T)+\Pi_{\rm power}(T), \quad 
\nonumber\\
\Pi_{\rm pert}(T)&=&
\left(\frac{m}{2\pi T}\right)^{3/2}
\left[1+\sqrt{2\pi mT}\alpha+\frac{1}{3}m\pi^2 T \alpha^2\right], \quad 
\nonumber\\
\Pi_{\rm power}(T)&=&
\left(\frac{m}{2\pi T}\right)^{3/2}
\left[-\frac{1}{4}\omega^2 T^2\left(1+\frac{11}{12}\sqrt{2\pi m T}\alpha\right)
+\frac{19}{480}\omega^4 T^4\right].
\end{eqnarray}
One can see here a ``perturbative contribution'' (i.e. the one not containing the confining potential),
and ``power corrections'' given in terms of the confining potential (including also mixed 
terms containing contributions from both Coulomb and confining potentials). 
A perturbative contribution may be written in the form of spectral representation \cite{voloshin} yielding
\begin{eqnarray}
\Pi_{\rm OPE}(T)&=&\int\limits_0^\infty dz e^{- z T}\rho_{\rm pert}(z)+\Pi_{\rm power}(T),
\nonumber\\
&&\rho_{\rm pert}(z)=
\left(\frac{m}{2\pi}\right)^{3/2}\left[2\sqrt{\frac{z}{\pi}}+\sqrt{2\pi m}\alpha+\frac{\pi^{3/2} m\alpha^2}{3\sqrt{z}}\right]
\end{eqnarray}
The ``physical'' representation for $\Pi(T)$---in the basis of hadron eigenstates---reads: 
\begin{eqnarray}
\Pi_{\rm phys}(T)=\langle \vec r_f=0|\exp(-H T)|\vec r_i=0\rangle = \sum_{n=0}^\infty R_n \exp(-E_n T), 
\quad R_n=|\Psi_n(\vec r=0)|^2.
\end{eqnarray}
Sum rule is just the statement that the correlator may be calculated in two ways---using  
the basis of quark states (OPE) or confined bound states---leading to the same result: 
\begin{eqnarray}
\Pi_{\rm OPE}(T)=\Pi_{\rm phys}(T). 
\end{eqnarray}
In order to isolate the ground-state contribution one needs the information about the excited states. 
A standard Ansatz for the hadron spectral density has the form \cite{svz} 
\begin{eqnarray}
\label{ansatz}
\rho_{\rm phys}(z)=R_g\delta(z-E_g)+\theta(z-z_{\rm eff})\rho_{\rm pert}(z).
\end{eqnarray}
It assumes that the contribution of the excited states may be described by contributions of diagrams of 
perturbation theory above some effective continuum threshold $z_{\rm eff}$. This effective continuum threhsold  
(different from the physical continuum threhsold which is determined by hadron masses) is an additional parameter of the 
method of sum rules. Using Eq.~(\ref{ansatz}) yields 
\begin{eqnarray}
R_g e^{- E_g T}=\int\limits_0^{z_{\rm eff}}dz e^{- z T}\rho_{\rm pert}(z)+\Pi_{\rm power}(T)\equiv \Pi_{\rm dual}(T,z_{\rm eff}).
\end{eqnarray}
As soon as one knows $z_{\rm eff}$, one immediately obtains estimates for $R_g$ and $E_g$, 
$R_{\rm dual}(T,z_{\rm eff})$ and $E_{\rm dual}(T,z_{\rm eff})=-d_T \log \Pi(T,z_{\rm eff})$. 
These however depend on unphysical parameters $T$ and $z_{\rm eff}$. 

\section{Exact effective continuum threshold}
Eq.~(\ref{ansatz}) is motivated by quark-hadron duality which claims that far above 
the threshold the hadron spectral density is 
well described by diagrams of perturbation theory. However, near the physical threshold---and this very region turns 
out to be essentail for the calculation of ground-state properties---the duality relation is violated. 
\begin{figure}[!ht]
\begin{center}
\begin{tabular}{c}
\includegraphics[width=8cm]{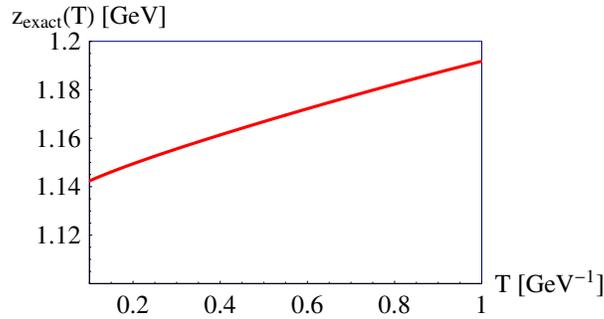}\\
\end{tabular}
\caption{\label{Plot:exactzeff}
Exact effective threshold for the polarization operator in a quantum-mechanical model for the 
potential containing confining harmonic-oscillator and Coulomb interactions (\protect\ref{Plot:exactzeff}).}
\end{center}
\end{figure}

\begin{figure}[!ht]
\begin{center}
\begin{tabular}{c}
\includegraphics[width=9cm]{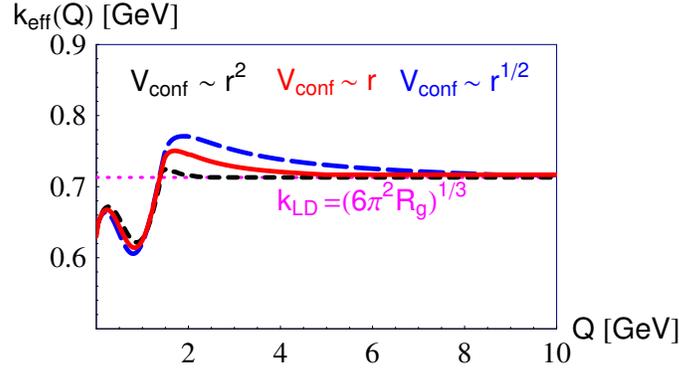}
\end{tabular}
\caption{\label{zeffLD1}
Exact effective threshold $k_{\rm eff}(T=0,Q)$ for the {\it elastic} form factor for various confining interactions 
$V_{\rm conf}(r)=\sigma_n\,(m\,r)^n$, $n=2,1,1/2$ \cite{irina}
Short-dashed (black) line: harmonic-oscillator confinement,
$n=2;$ full (red) line: linear confinement,~$n=1;$ long-dashed
(blue) line: square-root confinement, $n=1/2$. $R_{\rm g}\equiv |\Psi(r=0)|^2$. }
\end{center}
\end{figure}

\begin{figure}[!h]
\begin{center}
\begin{tabular}{c}
\includegraphics[width=9cm]{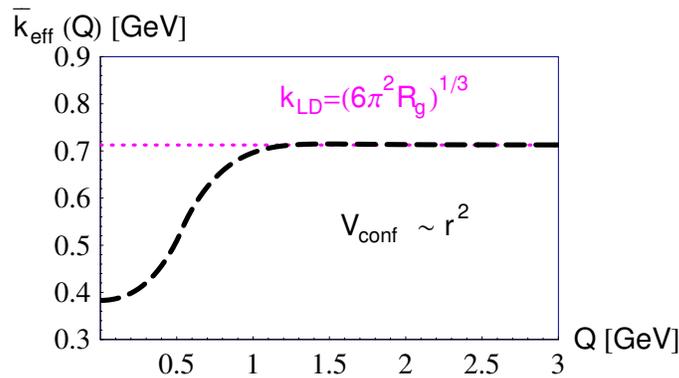}
\end{tabular}
\caption{\label{zeffLD2}
Exact effective threshold $k_{\rm eff}(T=0,Q)$ for the {\it transition} form factor 
for the case of harmonic-oscillator potential \cite{irina}. $R_{\rm g}\equiv |\Psi(r=0)|^2$.}
\end{center}
\end{figure}

The advantage of quantum mechanics is that the exact $E_g$ and $R_g$ may be obtained by solving Schr\"odinger 
equation \cite{schoberl}. We can then calculate $z_{\rm eff}$ from 
\begin{eqnarray}
R_g e^{- E_g T}=\int\limits_0^{z_{\rm eff}}dz e^{- z T}\rho_{\rm pert}(z)+\Pi_{\rm power}(T). 
\end{eqnarray}
Figures ~\ref{Plot:exactzeff},\ref{zeffLD1} and \ref{zeffLD2} provide the exact results for the effective thresholds 
obtained for the ground state of the Hamiltonian (\ref{conf+coulomb}). We consider a set of confining potentials 
$V_{\rm conf}(r)=\sigma_n\,(m\,r)^n$, $n=2,1,1/2$, and adopt parameter values appropriate for hadron physics, i.e.,
$m=0.175$ GeV for the reduced constituent light-quark mass and $\alpha=0.3$, 
and adapt the strengths $\sigma_n$ in our confining interactions such that the Schr\"odinger equation yields for each
potential the same $\Psi(r=0)=0.078$ GeV$^{3/2},$ which requires $\sigma_2=0.71$ GeV, $\sigma_1=0.96$ GeV, and
$\sigma_{1/2}=1.4$~GeV \cite{irina}. 

As shown in Fig.~\ref{Plot:exactzeff}, the obtained ``exact threshold'' $z_{\rm eff}(T)$ for the polarization operator 
turns out to be a slightly rising function of $T$ \cite{lms_2ptsr}. 


It should be taken into account that in general the effective continuum threhsold depends on the type 
of the correlator considered, on the details of the confining interaction, and for the three-point functions, 
also on $Q^2$. Figures \ref{zeffLD1} and \ref{zeffLD2} present the results of our calculations 
of the exact effective thresholds 
for the vacuum three-point correlators at $T=0$. Obviously, the exact effective thresholds 
in the region of small momentum transfers $Q$ depend rather sizeably on $Q^2$ and on the type of the hadron observable 
under consideration. Interestingly, due to the factorization properties of the amplitudes of exclusive processes at 
large momentum transfers, 
the exact effective thresholds for different correlators approach the same universal constant, related to the 
value of the bound-state wave function at the origin. For further details consult \cite{irina}.

\section{OPE and sum rule in QCD} 
Now, we briefly recall the main ingredients of the QCD sum-rule calculation of the ground-state decay constant. 
The basic object is the two-point function in QCD
\begin{eqnarray}
\label{Pi}
\Pi(p^2)=i \int dx e^{ipx}\langle \Omega|T\left(j_5(x)j_5^\dagger(0)\right)| \Omega \rangle, 
\qquad j_5(x)=(m_Q+m)\bar q i\gamma_5 Q(x)
\end{eqnarray} 
The physical QCD vacuum $|\Omega\rangle$ is expected to have a complicated nontrivial structure and to differ from 
perturbative QCD vacuum $|0\rangle$. One calculates (\ref{Pi}) by constructing the Wilsonian operator product 
expansion (OPE) for $T$-product of two current operators: 
\begin{eqnarray}
T\left(j_5(x)j_5^\dagger(0)\right)=C_0(x^2,\mu)\hat 1 + \sum\limits_n C_n(x^2,\mu) :\hat O(0,\mu):
\end{eqnarray}
One now needs to describe properties of the physical QCD vacuum. This is done by introducing condensates  -- 
nonzero expectation values of gauge-invariant operators over physical vacuum: 
\begin{eqnarray}
\langle \Omega | :\hat O(0,\mu):| \Omega\rangle\ne 0.
\end{eqnarray}
As the next step, one obtains the dispersion representation for the purely perturbative contribution to $\Pi(p^2)$ and 
performs the Borel transform ($p^2\to \tau$) which corresponds to turning from Green functions in Minkowski space 
to time-evolution operator in Euclidean space. The Borel transform kills the possible subtraction terms in the 
dispersion representation for $\Pi(p^2)$ and improves the convergence of the perturbative expansion. 
Finally, one comes to 
\begin{eqnarray}
\Pi(\tau)&=&
\int\limits^{\infty}_{(m_Q+m_u)^2} e^{-s\tau}\rho_{\rm pert}(s,\alpha,m_Q,\mu)\,ds + \Pi_{\rm power}(\tau,m_Q,\mu),
\nonumber\\
\rho_{\rm pert}(s,\mu)&=&\rho^{(0)}(s)+\frac{\alpha_s(\mu)}{\pi}\rho^{(1)}(s)+
\left(\frac{\alpha_s(\mu)}{\pi}\right)^2\rho^{(2)}(s)+\cdots
\end{eqnarray}
where $\Pi_{\rm power}(\tau,\mu)$ is the nonperturbative contribution to $\Pi(\tau)$. $\Pi_{\rm power}(\tau,\mu)$ is 
given by a power expansion in $\tau$ in terms of the condensates and rad corrections to them. 
The rum rule reads 
\begin{eqnarray}
\Pi_{\rm OPE}(\tau)=\Pi_{\rm hadron}(\tau)
\end{eqnarray}
The hadronic part of the sum rule contains the contributions of the ground state and the hadron-continuum states. 
At this point one invokes quark-hadron duality approximation: 
\begin{eqnarray}
\int^\infty_{s_{\rm eff}} ds \exp(-s \tau)\rho_{\rm pert}(s)=\int^\infty_{s_{\rm phys.cont.}} ds \exp(-s \tau)\rho_{\rm hadr}(s). 
\end{eqnarray}
With the help of this duality approximation, the contribution of the excited states cancels against the 
high-energy region of the perturbative contribution, and relates---quite similar to the case of quantum mechanics---
the ground-state contribution to the low-energy part of Feynman diagrams of perturbative QCD and power corrections 
\cite{jamin}: 
\begin{eqnarray}
\label{SR_QCD}
f_Q^2 M_Q^4 e^{-M_Q^2\tau}=
\int\limits^{s_{\rm eff}}_{(m_Q+m_u)^2} e^{-s\tau}\rho_{\rm pert}(s,\alpha,m_Q,\mu)\,ds + 
\Pi_{\rm power}(\tau,m_Q,\mu)\equiv \Pi_{\rm dual}(\tau,\mu,s_{\rm eff})
\end{eqnarray}
In order to extract the decay constant one should fix the effective continuum threshold $s_{\rm eff}$ which,
as illustrated in quantum mechanics, should be a function of $\tau$;  
otherwise the $\tau$-dependences of the l.h.s. and the r.h.s. of (\ref{SR_QCD}) do not match each other. 
The exact $s_{\rm eff}$ corresponding to the exact hadron decay constant 
and mass on the l.h.s. is of course not known. The extraction of hadron parameters from the 
sum rule consists therefore in attempting 
(i) to find a good approximation to the exact threshold and 
(ii) to control the accuracy of this approximation. 
For further use, we define the dual decay constant $f_{\rm dual}$ and the dual invariant mass $M_{\rm dual}$ by relations
\begin{eqnarray}
\label{fdual}
f_{\rm dual}^2(\tau)=M_Q^{-4} e^{M_Q^2\tau}\Pi_{\rm dual}(\tau, s_{\rm eff}(\tau)),\\\quad 
M_{\rm dual}^2(\tau)=-\frac{d}{d\tau}\log \Pi_{\rm dual}(\tau, s_{\rm eff}(\tau)). 
\end{eqnarray}

\section{A new algorithm for fixing the effective continuum threshold}
If the mass of the ground state is known, the deviation of the dual mass from the actual mass of the ground state 
gives an indication of the contamination of the dual correlator by excited states. 
Assuming a specific functional form of the effective threshold and requiring the least deviation of the dual mass 
(\ref{fdual}) from the known ground-state mass in the $\tau$-window leads to a variational solution for the 
effective threshold. As soon as the latter has been fixed, one calculates the decay constant from (\ref{fdual}). 

{\bf Our algorithm for extracting ground-state parameters reads:}
\begin{itemize}
\item[(i)]
Consider a set of Polynomial $\tau$-dependent Ansaetze for $s_{\rm eff}$: 
\begin{eqnarray}
s^{(n)}_{\rm eff}(\tau)=s^{(n)}_0+s^{(n)}_1\tau+s^{(n)}_2 \tau^2+\ldots.
\end{eqnarray}
\item[(ii)]
Calculate the dual mass for these $\tau$-dependent thresholds and minimize the squared difference 
between the ``dual'' mass $M^2_{\rm dual}$ and the known value $M^2_B$ in the $\tau$-window
\begin{eqnarray}
\label{chisq}
\chi^2 \equiv \frac{1}{N} \sum_{i = 1}^{N} \left[ M^2_{\rm dual}(\tau_i) - M_Q^2 \right]^2.
\end{eqnarray}  
This gives the parameters of the effective continuum thresholds $s^{(n)}_i$. 
\item[(iii)]
Making use of the obtained thresholds, calculate the decay constant. 
\end{itemize}

\begin{figure}[!ht]
\begin{center}
\begin{tabular}{cc}
\includegraphics[width=6.5cm]{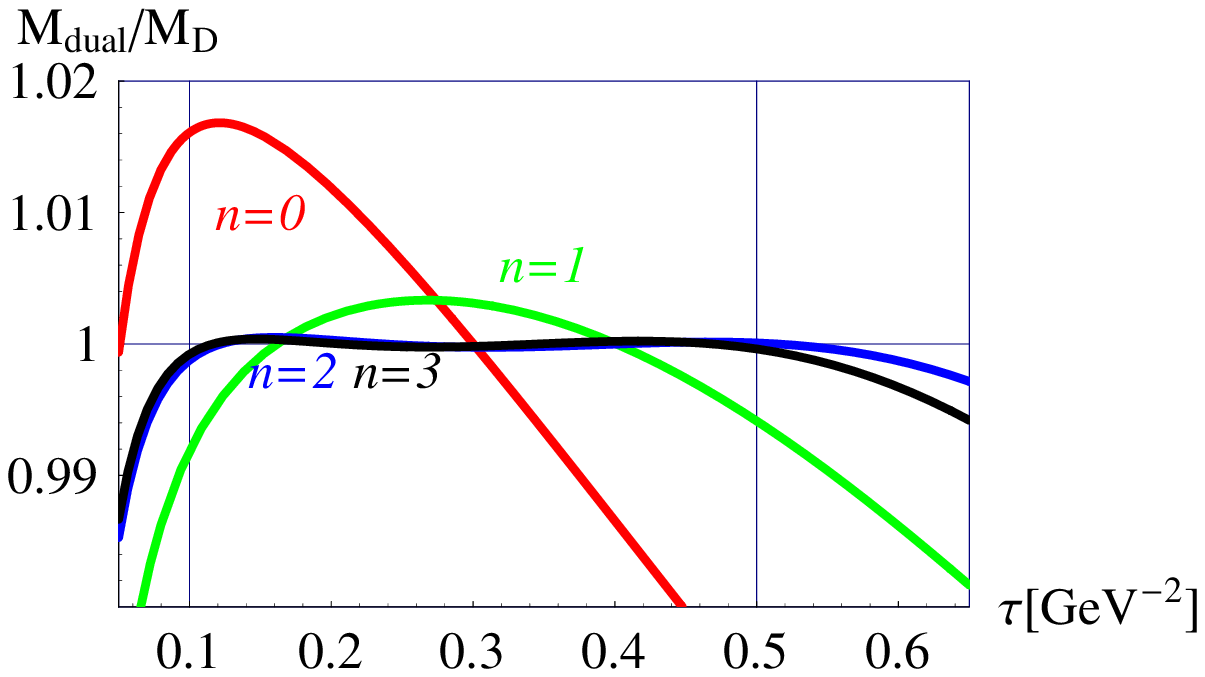}&\includegraphics[width=6.5cm]{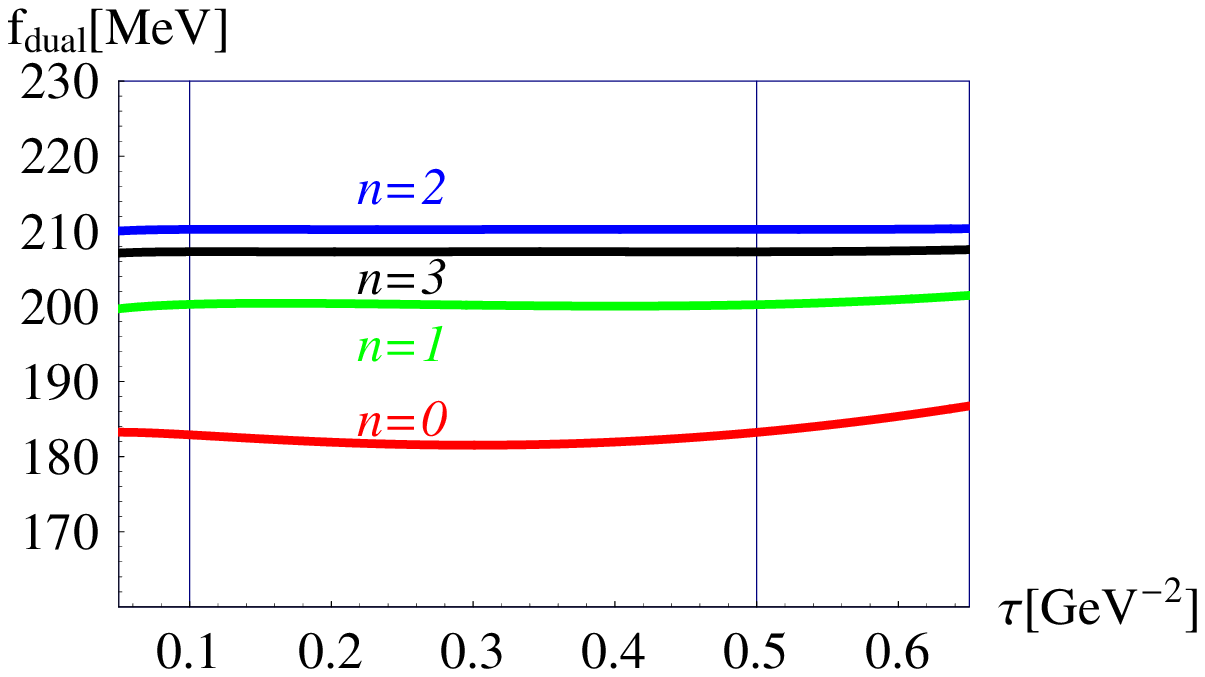}\\
\end{tabular}
\caption{\label{Fig:4}
The dual mass (left) and the dual decay constant (right) for different Ans\"atze for the effective continuum threshold.}
\end{center}
\end{figure}

Figure \ref{Fig:4} illustrates the application of our procedure of fixing the effective continuum 
threshold for the realistic 
case of the $D$-meson decay constant. Clearly, the $\tau$-dependent effective continuum threshold leads to a much 
better stability of the dual mass and allows one to reproduce the knows mass of the ground state in a rather broad 
window of the Borel parameter.

The crucial question which emerges at this point is how to interprete the results obtained with different 
effective thresholds. 
To answer this question we compare the extracion of the decay cosntant in QCD (where the exact result is not known) with the 
extraction of the decay constant in potential model (where the exact result is known). 
 
\begin{figure}[!b]
\begin{center}
\begin{tabular}{cc}
\includegraphics[width=6.5cm]{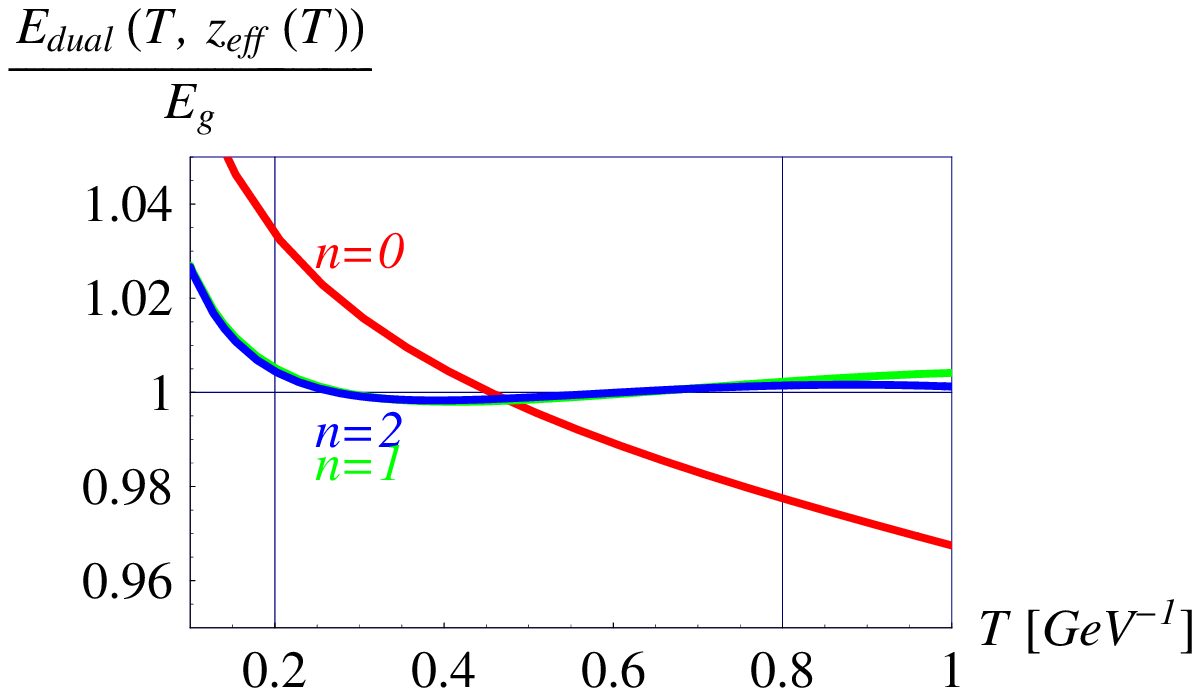}&\includegraphics[width=6.5cm]{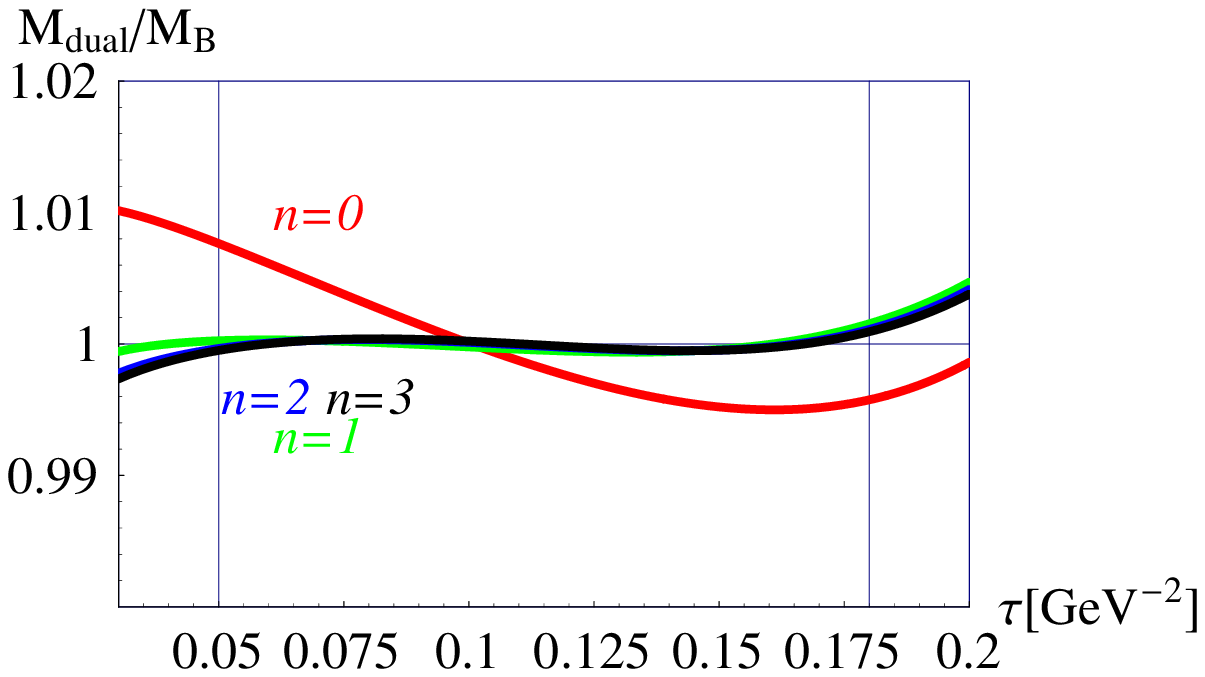}\\
\includegraphics[width=6.5cm]{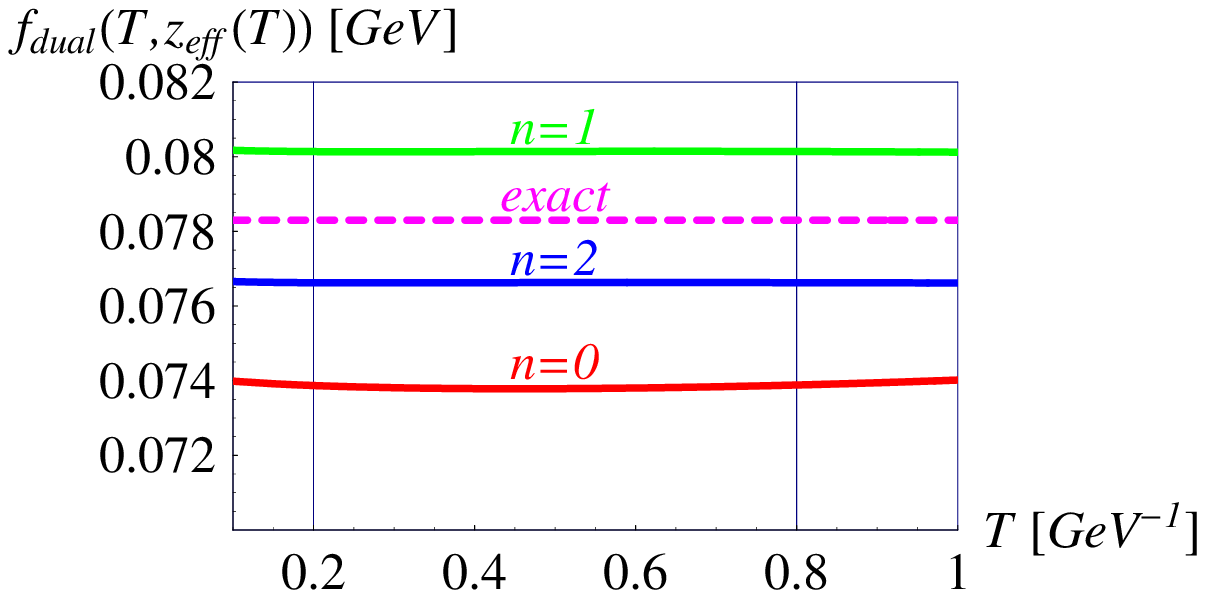}&\includegraphics[width=6.5cm]{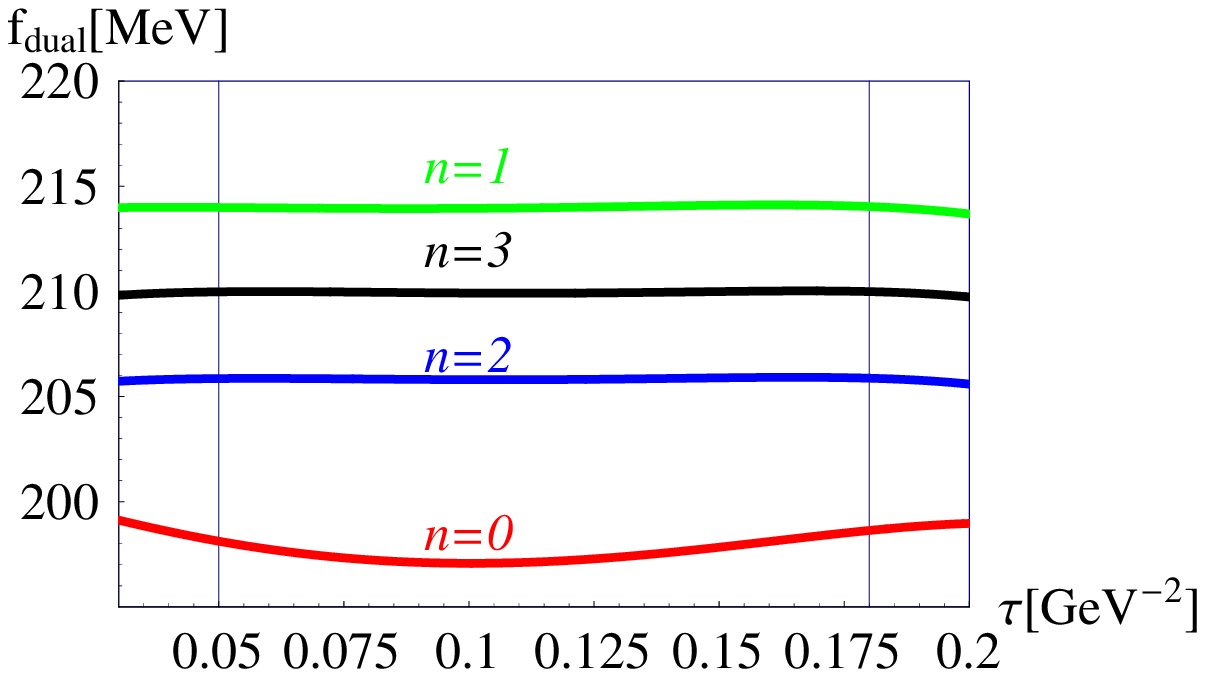}
\end{tabular}
\caption{\label{Fig:5}
The outcome of our algorithm in potential model (left) and in QCD (right).}
\end{center}
\end{figure}

Figure \ref{Fig:5} compares the application of our algorithm in quantum mechanics and in QCD \cite{lms_qcdvsqm}. 
In both cases a very similar pattern emerges. In potential model, the exact value of the decay constant lies in the band 
of values provided by the results corresponding to {linear}, {quadratic}, and {cubic} effective thresholds. 
We have checked that this result holds independently of the specific shape of the confining potential 
and for a broad range of the parameters of the potential model. Since the extraction procedures in potential 
model and in QCD exhibit precisely the same pattern, 
also in QCD we take this band of values as the estimate of the intrinsic uncertainty of the sum-rule estimate for 
the decay cosntant. So we come to the final point of our algorithm:
\begin{itemize}
\item[(iv)]
Take the band of values provided by the results corresponding to {\bf linear}, 
{\bf quadratic}, and {\bf cubic}
effective thresholds as the characteristic of the intrinsic uncertainty of the extraction procedure. 
\end{itemize}

Obviously, the $\tau$-dependent polynomial effective continuum threshold allows 
one to reproduce the known mass of the ground state much better; this means that the contrinution of the ground state 
from the dual correlator is isolated with a better accuracy. Consequently, the accuracy of the extraction of the 
ground-state decay constant imroves visibly. 

\section{Decay constants of charmed mesons}
We now discuss the application of our algorithm to the extraction of the decay cosntants of $D$ and $D_s$ mesons \cite{lms_fp}.

\subsection{The Borel window}
First, we must fix our working $\tau$-window where, on the one
hand, the OPE gives a sufficiently accurate description~of the
exact correlator (i.e., higher-order radiative and power
corrections are small) and, on the other hand, the ground state
gives a ``sizable'' contribution to the correlator. Since the
radiative corrections to the condensates increase rather fast with
$\tau$, it is preferable to stay at the lowest possible values of
$\tau$. We shall therefore fix the window by the following
criteria \cite{lms_fp}: (a) In the window, power
corrections $\Pi_{\rm power}(\tau)$ should not exceed 30\% of the
dual correlator $\Pi_{\rm dual}(\tau,s_0)$. This restricts the
upper boundary of the $\tau$-window. The ground-state contribution
to the correlator at this value of~$\tau$ comprises about 50\% of
the correlator. (b) The lower boundary of the $\tau$-window is
fixed by the requirement that the ground-state contribution does
not fall below 10\%.

\subsection{Uncertainties in the extracted decay constant}
Clearly, the extracted value of the decay constant is sensitive (i) to
the precise values of the OPE parameters and (ii) to the prescription
for fixing the effective continuum threshold: trying different Ans\"atze for the effective continuum
threshold, one obtains different estimates for the decay constant. 
The corresponding errors in the resulting decay constants~are called the 
{\it OPE-related error} and the {\it systematic error}, respectively.
Let us discuss these in more detail. 

\subsubsection{OPE-related error}
We perform the analysis making use of the $\overline{MS}$ scheme in which case the OPE exhibits a reasonable convergence. 
The corresponding OPE parameters are summarized here: 
\begin{eqnarray}
&&\overline{m}_c(\overline{m}_c)=(1.279\pm 0.013){\rm GeV}, m(2\;{\rm GeV})=(3.5\pm 0.5) {\rm MeV}, 
m_s(2\;{\rm GeV})=(100\pm 10) {\rm MeV}, \nonumber\\
&&\alpha_S(M_Z)=0.1176\pm 0.0020, \left\langle\frac{\alpha_s}{\pi}GG\right\rangle=(0.024\pm 0.012)\;{\rm GeV}^4\nonumber\\
&&\langle\bar qq\rangle(2\;{\rm GeV})=-((267\pm 17)\;{\rm MeV})^3, 
\langle\bar ss\rangle(2\;{\rm GeV})/\langle\bar qq\rangle(2\;{\rm GeV})=0.8\pm 0.3.
\end{eqnarray}
The value of the OPE-related error is obtained as follows: We
perform a bootstrap analysis by allowing the OPE
parameters to vary over their ranges using 1000 bootstrap events. 
Gaussian distributions for all OPE parameters but $\mu$ are employed. 
For $\mu$ we assume a uniform distribution in the corresponding range, which~we~choose 
to be $1$ GeV $ \leq \mu \leq 3$ GeV for charmed mesons. The resulting
distribution of the decay constant turns out to be close to Gaussian shape. 
Therefore, the quoted OPE-related error is a Gaussian~error.

\subsubsection{Systematic error}
The systematic error of a hadron parameter determined by the method of sum rules 
(i.e., the error related to the limited accuracy of this method) represents perhaps the most
subtle point in the applications of this method. So far no way to arrive at a {\it rigorous}---in the mathematical
sense---systematic error has been proposed. Therefore, we have to rely on our experience obtained from the
examples where the exact hadron parameters may be calculated independently of the method of dispersive sum rules and then
compared with the results of the sum-rule approach. As we have seen above, the band of values obtained from linear, 
quadratic,~and 
cubic Ans\"atze for the effective threshold encompasses the true value of the decay constant; 
moreover, the extraction procedures in quantum mechanics and in QCD are even quantitatively rather similar. 
Therefore, we believe that the half-width of this band may be regarded as {\it realistic\/} estimate for the 
systematic uncertainty of the decay constant. Presently, we do not see other possibilities to obtain a
more reliable estimate for the systematic~error.

\subsection{Decay constant of the $D$ meson}
The $\tau$-window for the charmed mesons, $\tau=(0.1 - 0.5)\;\mbox{GeV}^{-2}$, 
is chosen according to the criteria formulated~above. 
\begin{figure}[!b]
\begin{center}
\begin{tabular}{cc}
\includegraphics[width=6cm]{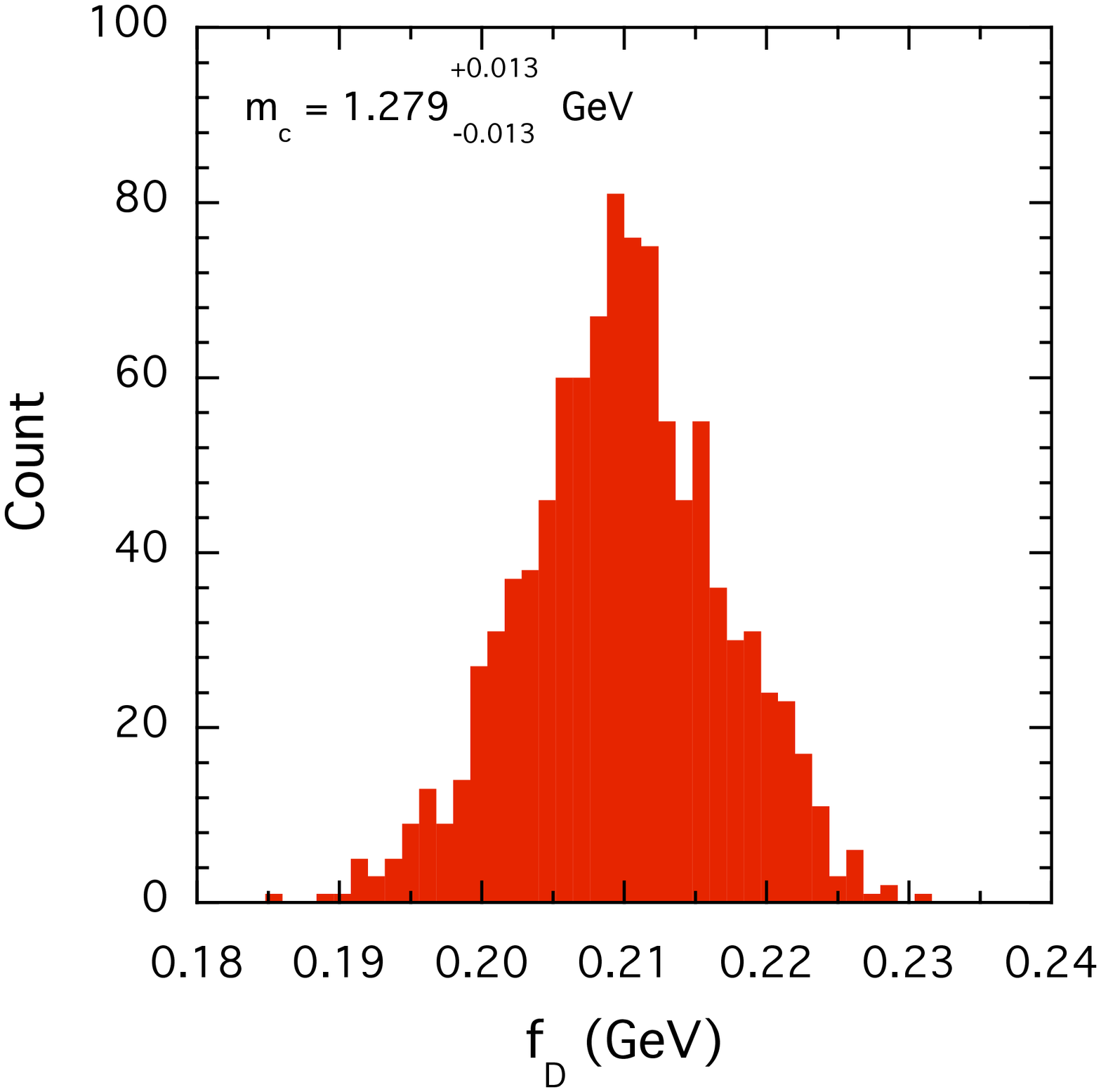}\qquad&
\includegraphics[width=6cm]{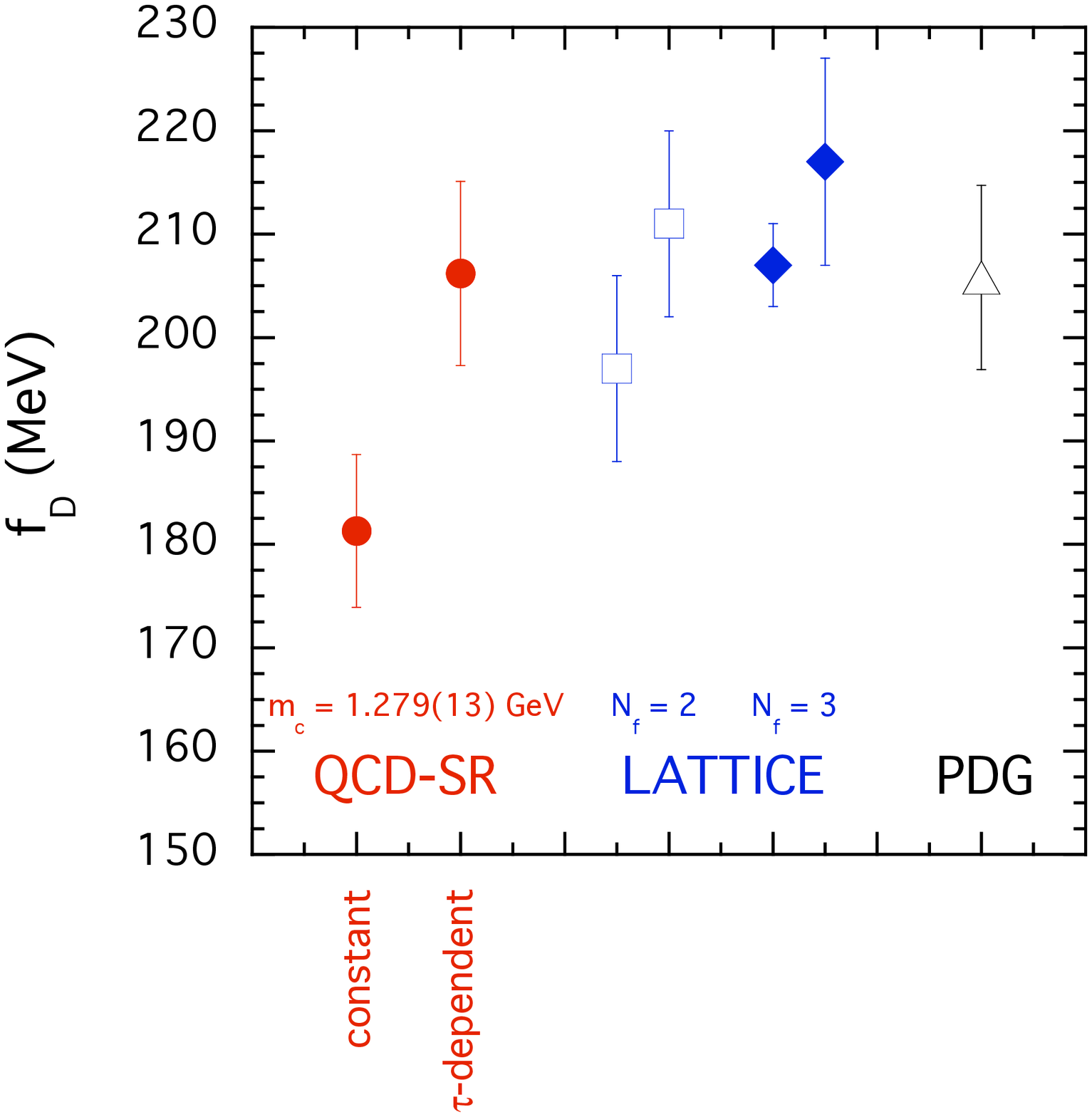}
\end{tabular}
\caption{\label{Plot:fD_bootstrap} 
(a) Distribution of $f_D$
obtained by the bootstrap analysis of the OPE uncertainties.
(b) Summary of findings for $f_D$. 
The estimate obtained with the constant threshold includes the OPE uncertainty only; 
for the $\tau$-dependent QCD-SR result the error shown is the sum of the OPE and systematic
uncertainties in (\protect\ref{fD}), added in quadrature.}
\end{center}
\end{figure}
Figure~\ref{Fig:4} shows the application of
our procedure of fixing the effective continuum threshold and extracting the resulting $f_D$. 
Figure~\ref{Plot:fD_bootstrap}a 
depicts the result of the bootstrap analysis of the OPE uncertainties. The distribution has
a Gaussian shape, and therefore the corresponding OPE uncertainty
is the Gaussian error. Adding the half-width of the band deduced
from our $\tau$-dependent Ans\"atze~for the effective continuum
threshold of degree $n=1,2,3$ as the (intrinsic) systematic error,
we obtain the following~result:
\begin{equation}
\label{fD} f_{D}= \left(206.2 \pm 7.3_{\rm (OPE)} \pm 5.1_{\rm (syst)}\right) \mbox{MeV}.
\end{equation}
The main sources of the OPE uncertainty in the extracted $f_D$ are
its renormalization-scale dependence and the error~of the quark
condensate.

For a $\tau$-independent Ansatz for the effective continuum
threshold a bootstrap analysis entails the substantially~lower
range
$f_D^{(n=0)} = \left(181.3\pm 7.4_{\rm (OPE)}\right) \mbox{MeV}$,
which differs from our $\tau$-dependent result (\ref{fD}) by
$\simeq 10\%$, i.e.,~by almost~three times the OPE uncertainty.
Moreover, as we have shown, making use of merely the constant Ansatz for the
effective continuum threshold does not allow one to probe at all
the intrinsic systematic error of the QCD sum rule. From our
result (\ref{fD}) the latter turns out to be of the same order as
the OPE uncertainty.

Allowing the threshold to depend on $\tau$ leads to a clearly
visible effect and brings the results from QCD sum~rules~into
perfect agreement with recent lattice results and the experimental
data (Fig.~\ref{Plot:fD_bootstrap}b). This perfect agreement of
our~result with both experimental data and lattice results
provides a strong argument in favour of the reliability of our
procedure.

\subsection{Decay constant of $D_s$ meson}
We now apply our algorithm to the decay cosntant of $D_s$ mesons. 
\begin{figure}[!h]
\begin{center}
\begin{tabular}{cc}
\includegraphics[width=6cm]{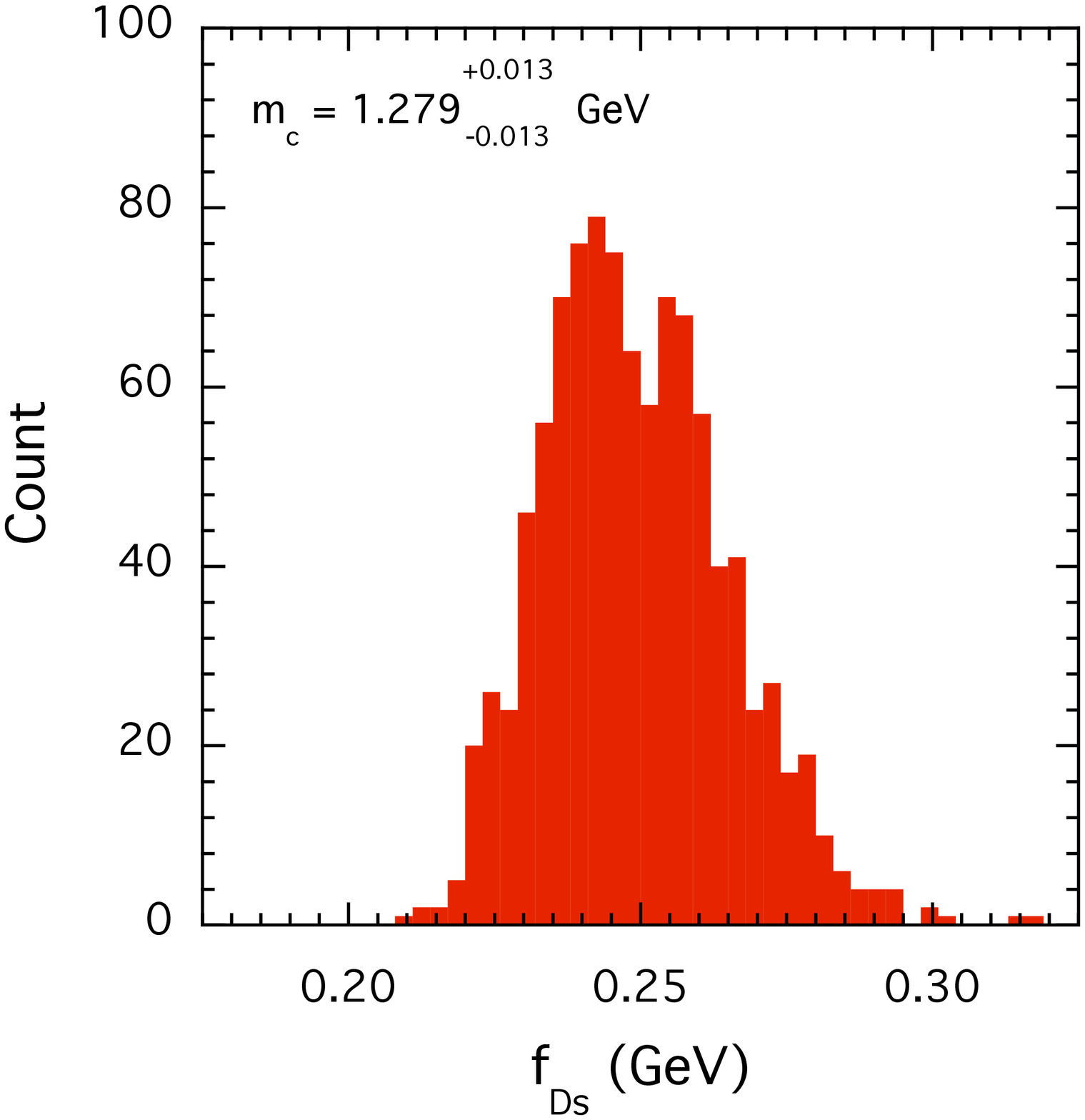}\qquad&
\includegraphics[width=6cm]{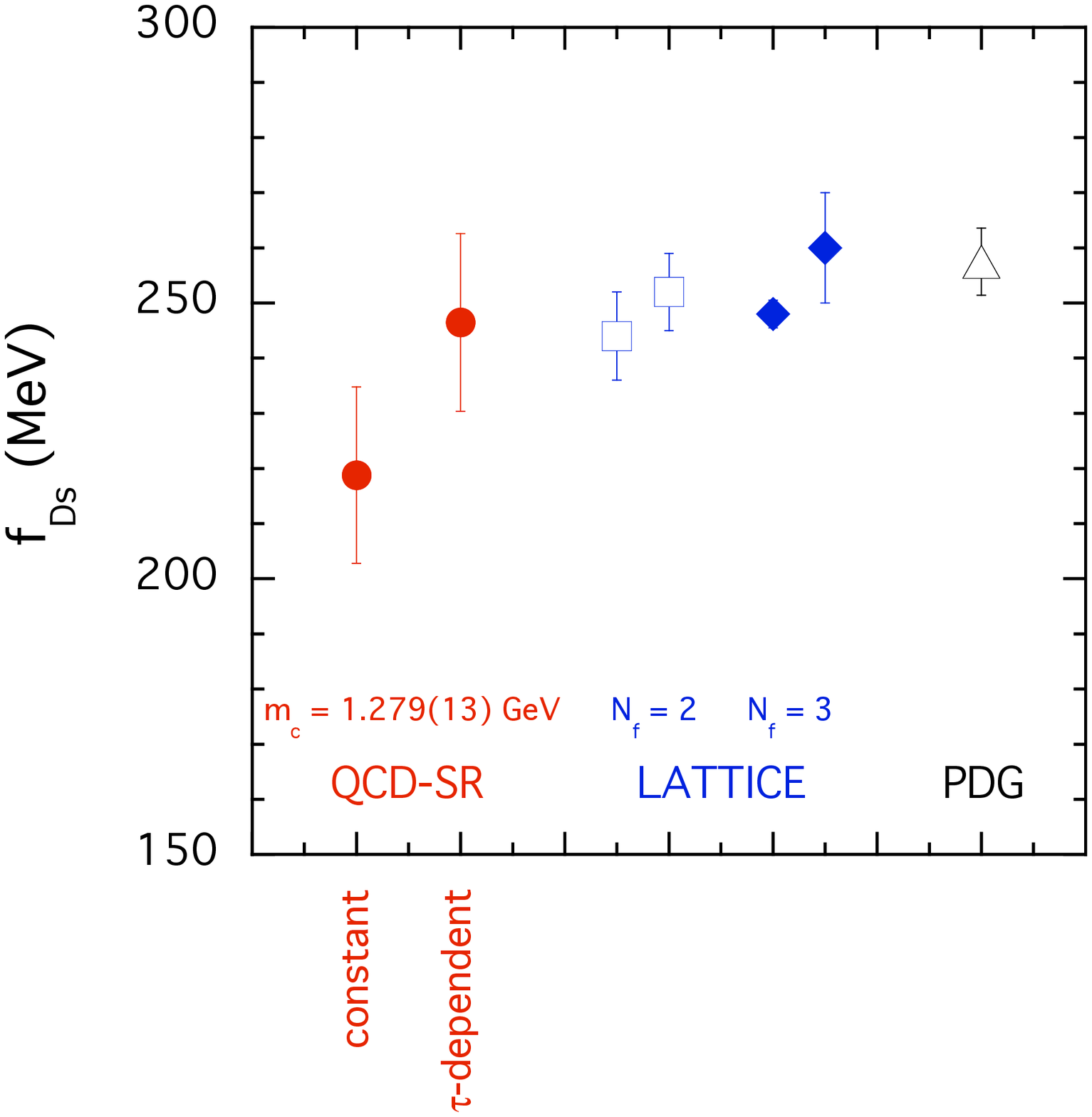}
\end{tabular}
\caption{\label{Plot:fDs_bootstrap} 
(a) Distribution of $f_{D_s}$
obtained by the bootstrap analysis of the OPE uncertainties.
Gaussian distributions for all~OPE parameters but $\mu$. For
$\mu$ we assume a uniform distribution in the~range 
$1\;{\rm GeV} < \mu < 3\;{\rm GeV}$. 
(b) Summary of findings for $f_{D_s}$.
Lattice results are from \cite{ETMC1,ETMC2} for two dynamical
light~flavors ($N_f = 2$) and from \cite{HPQCD1,FNAL+MILC1} for
three dynamical flavors ($N_f=3$). The triangle represents the
experimental value from PDG \cite{pdg}. The estimate obtained with 
the constant threshold includes the OPE uncertainty only; for the $\tau$-dependent
QCD-SR result the error shown is the sum of the OPE and systematic
uncertainties in (\protect\ref{fDs}), added in quadrature.}
\end{center}
\end{figure}
Performing the bootstrap analysis of the~OPE uncertainties, we
obtain the following estimate:
\begin{eqnarray}
\label{fDs} 
f_{D_s} = \left(245.3 \pm 15.7_{\rm (OPE)} \pm 4.5_{\rm (syst)}\right){\rm MeV}.
\end{eqnarray}

As in the case of $f_D$, a constant-threshold Ansatz yields a
substantially lower value: $f_{D_s}^{(n=0)} = \left(218.8 \pm
16.1_{\rm (OPE)}\right){\rm MeV}$.

\subsection{The ratio $f_{D_s}/f_D$}
For the ratio of the $D$ and $D_s$ decay constants we report the
sum-rule prediction
\begin{eqnarray}
\label{ratioD}
f_{D_s}/f_D = 1.193\pm 0.025_{(\rm OPE)}\pm 0.007_{(\rm syst)}.
\end{eqnarray}
This value is to be compared with the PDG average $f_{D_s}/f_D =1.25\pm 0.06$ \cite{pdg} 
as well as with the recent lattice results $f_{D_s}/f_D = 1.24 \pm 0.03$ \cite{ETMC1} for $N_f=2$ and
$f_{D_s}/f_D = 1.164 \pm 0.011$ \cite{HPQCD1} and $f_{D_s}/f_D =1.20 \pm 0.02$ \cite{FNAL+MILC1} for $N_f=3$. 
The error in (\ref{ratioD}) arises mainly from the uncertainties in the quark
condensates $\langle \bar s s\rangle/\langle \bar q q\rangle=0.8\pm 0.3$.

\subsection{Conclusions on decay constants of charmed mesons}
We presented a detailed analysis of the decay constants of charmed
heavy mesons with the help of QCD sum rules. Particular emphasis
was laid on the study of the uncertainties in the extracted values
of the decay constants: the~OPE uncertainty related to the not
precisely known QCD parameters and the intrinsic uncertainty of
the sum-rule method related to a limited accuracy of the extraction procedure.

Our main findings may be summarized as follows.

\vspace{.2cm}
\noindent
(i) The perturbative expansion of the two-point function in terms
of the pole mass of the heavy quark exhibits no sign of
convergence. However, reorganizing this expansion in terms of the
corresponding running mass leads to~a clear hierarchy of the
perturbative contributions. Interestingly, the decay constant
extracted from the pole-mass~OPE proves to be sizeably smaller
than the one extracted from the running-mass OPE. In spite of this
numerical difference, the decay constants extracted from these two
correlators exhibit perfect stability in the Borel parameter \cite{lms_fp}. This
example clearly demonstrates that stability {\it per se\/} does not guarantee the
reliability of the sum-rule extraction of any bound-state
parameter.

\vspace{.2cm}
\noindent
(ii) We have made use of the Borel-parameter-dependent effective
threshold for the extraction of the decay constants. The
$\tau$-dependence of the effective threshold emerges quite
naturally when one attempts to increase the accuracy of the
duality approximation. According to our algorithm, one should
consider different polynomial Ans\"atze for the~effective
threshold and fix the coefficients in these Ans\"atze by
minimizing the deviation of the dual mass from the known~actual
meson mass in the window. Then, the band of values corresponding
to the linear, quadratic, and cubic Ans\"atze reflects the
intrinsic uncertainty of the method of sum rules. The efficiency
of this criterion has been tested before for several examples of
quantum-mechanical models. This strategy has now been applied to
the decay constants of heavy mesons.

\vspace{.2cm}
\noindent
(iii) We obtained the following sum-rule estimates for the decay
constants of the charmed $D$ and $D_s$ mesons:
\begin{eqnarray}
\label{fD_final} f_{D}&=& \left(206.2 \pm 7.3_{\rm (OPE)} \pm 5.1_{\rm (syst)}\right) \mbox{MeV}, \\ 
\label{fDs_final} f_{D_s}&=& \left(245.3 \pm 15.7_{\rm (OPE)} \pm 4.5_{\rm (syst)}\right) \mbox{MeV}.
\end{eqnarray}
We point out that we provide both the OPE uncertainties and the
intrinsic (systematic) uncertainty of the method~of sum rules
related to the limited accuracy of the extraction procedure. In
the case of $f_D$ the latter turns out to be~of~the same order as
the OPE uncertainty. Noteworthy, adopting a $\tau$-independent
effective threshold leads to a substantially lower range
$f_D^{(n=0)} = \left(181.3 \pm 7.4_{\rm (OPE)}\right) \mbox{MeV}$, which differs from our $\tau$-dependent result (\ref{fD_final})
by almost~three~times the OPE uncertainty. 

\vspace{.2cm}
\noindent
(iv) Our study of charmed mesons clearly demonstrates that the use
of Borel-parameter-dependent thresholds leads to two essential
improvements:

a. The actual accuracy of the decay constants extracted from sum
rules improves considerably.

b. Our algorithm yields {\it realistic\/} (although not entirely
rigorous) estimates for the systematic errors and allows~one~to
reduce their values to the level of a few percent. Due to the
application of our prescription, the QCD sum-rule results are
brought into perfect agreement both with the experimental results
and with lattice QCD.


\section{Conclusions}
The effective continuum threshold $s_{\rm eff}$ is an important ingredient of the method of dispersive 
sum rules which determines to a large extent the numerical values of the extracted hadron 
parameter. Finding a criterion for fixing $s_{\rm eff}$ poses a problem in the method of sum rules. 

\vspace{.2cm}
\noindent 
$\bullet$ $s_{\rm eff}$ depends on the {\it external kinematical variables} (e.g., 
momentum transfer in sum rules for 3-point correlators and light-cone sum rules) and 
{`\it `unphysical'' parameters} (renormalization scale $\mu$, Borel parameter $\tau$). 
Borel-parameter $\tau$-dependence of $s_{\rm eff}$ emerges naturally when trying to make 
quark-hadron duality more accurate. 

\vspace{.2cm}\noindent 
$\bullet$ We proposed a new algorithm for fixing $\tau$-dependent $s_{\rm eff}$, for those problems 
where the bound-state mass is known. We have tested that our algorithm leads to more accurate 
values of ground-state parameters than the ``standard'' algorithms used in the context 
of dispersive sum rules before. Moreover, our algorithm allows one to probe 
{\it ``intrinsic''} uncertainties related to the limited accuracy of 
the extraction procedure in the method of QCD sum rules.  
\vspace{.5cm}

\noindent 
\acknowledgments 
I have pleasure to thank Wolfgang Lucha and Silvano Simula for the most pleasant and fruitful collaboration 
on the subject presented in this talk. The work was supported by the Austrian Science Fund FWF under Project P22843, 
by a grant for leading scientific schools 1456.2008.2, and by FASI state Contract No. 02.740.11.0244.

\end{document}